\documentclass[10pt,a4paper]{revtex4}

\usepackage{amsmath}
\usepackage{amsthm}
\usepackage{amssymb}
\usepackage[a4paper]{geometry}

\usepackage{graphicx}
\usepackage{tkz-berge}		
\usepackage{pgfplotstable}
\usepackage{color}

\usepackage[linesnumbered]{algorithm2e} 	

\DeclareMathOperator*{\argmax}{arg\,max}

\begin{document}

\author{
Guido Previde Massara$^1$,
T. Di Matteo$^2$
Tomaso Aste$^{1,3}$}
\affiliation{
$^1$ Department of Computer Science, University College London,  Gower Street, London, WC1E 6BT, UK\\ 
$^2$ Department of Mathematics, King's College London,  The Strand, London, WC2R 2LS, UK\\
$^3$ Systemic Risk Centre, London School of Economics and Political Sciences, London, WC2A2AE, UK}

\date{16/07/15}

\title{ 
Network Filtering for Big Data: \\ Triangulated Maximally Filtered Graph
}

\begin{abstract}
We propose a  network-filtering method, the Triangulated Maximally Filtered Graph (TMFG), that provides an approximate solution to the \textsc{Weighted Maximal Planar Graph} problem.
The underlying idea of TMFG consists in building a triangulation that maximizes a score function associated with the amount of information retained by the network. 
TMFG uses as weights any arbitrary similarity measure to arrange data into a meaningful network structure that can be used for clustering, community detection and modeling. 
The method is fast, adaptable and scalable to very large datasets, it allows online updating and learning as new data can be inserted and deleted with combinations of local and non-local moves. 
TMFG permits readjustments of the network in consequence of changes in the strength of the similarity measure. 
The method is based on local topological moves and can therefore take advantage of parallel and GPUs computing. 
We discuss how this network-filtering method can be used intuitively and efficiently for big data studies and its significance from an information-theoretic perspective.
\\


\end{abstract}

\maketitle

\section{Introduction}
We are witnessing interesting times rich of information, readily available for us all.
Using, understanding and filtering such information has become a major activity across science, industry and society at large.
Our society has become a global information processing system where news propagate and  impact on individuals and the economy at increasingly fast rates with increasingly large effects.
It is therefore important to have tools that can analyse this information while it is generated and that can provide ways to reduce complexity and dimensionality while keeping the integrity of the dataset.    
Information content and flow are often associated with large degrees of redundancy both in time (repeating and scaling patterns) and across different variables (similarity, dependency and causality).
Redundancy is often used to convey strength to the meaning or, more simply, it is the signal of recurring patterns with high statistical significance and therefore important.
In this paper we propose to use such redundancy to build an information-based network that retains the relevant part of the data-interdependency structure. 
The structure of this network is a representation of the information in the dataset and such information can be efficiently analysed by using network-theoretic tools.

The idea of using redundancy -- mostly correlation coefficients --  to filter information in complex datasets by building sparse networks retaining relevant edges only has been very actively studied in the literature mostly by means of two approaches: 
i) the minimum spanning tree (MST) \cite{Prim57,Mantegna99}; 
ii) the planar maximally filtered graph (PMFG) \cite{aste2005complex,Tumminello26072005,MatlabPMFG}. 
The common idea underneath these two approaches is to filter a dense matrix of weights by retaining the largest and most significant possible subgraph while imposing global constraints on the topology of the resulting network.     
In particular, in the MST approach edges with the largest weights (e.g. correlations) are retained while constraining the  subgraph to be globally a (spanning) tree.
Similarly, in the PMFG construction the largest weights (e.g. largest correlation coefficients) are retained while constraining the subgraph to be globally a planar graph (see \cite{Mantegna99,aste2005complex,Tumminello26072005}). Both the MST and the PMFG are particular cases of simplicial complexes; our proposed method can be extended to more general simplicial complexes with different constraints (for instance on the topological genus, or on the size of the largest complete subgraph -- or \emph{clique}).

The PMFG is a greedy solution of the \textsc{Weighted Maximum Planar Graph} (WMPG) problem: given a complete edge-weighted graph find a planar subgraph which is maximal (i.e. no edge can be added without destroying planarity) and such that the sum of the edge weights is maximum. The problem is known to be NP-complete (see \cite{giffin1984graph} for a proof and \cite{Osman2003} for a review), but algorithms providing sub-optimal solutions are known. For instance, an approximation algorithm with a guaranteed performance ratio of \(\frac{3}{8}\) for complete graphs is discussed in \cite{Calinescu2003}. 

The PMFG has a richer information content than the MST with a larger number of edges (the PMFG has $3 p-6$ edges, while the MST has $p-1$, where $p$ is the number of vertices) and contains of 3- and 4-cliques.
However, the network is still sparse, filtering $3 p-6$ edges out of $p(p-1)/2$ of the complete graph $K_p$ which is associated with the original dense matrix of weights.

Planar filtered graphs are powerful tools to study complex datasets. 
It has been shown in \cite{song2012hierarchical} that by making use of the 3-clique structure of the PMFG a  clustering structure can be extracted allowing dimensionality reduction that keeps both local information and global hierarchy in a deterministic manner without the use of any prior information. 
Applications of Planar filtered graphs to financial data-sets can meaningfully identify industrial activities and structural market changes \cite{musmeci2014relation,musmeci2014risk}. 
Planar filtered graphs can be used to diversify financial risk by building a well-diversified portfolio that effectively reduces  risk by investing in stocks that occupy peripheral regions of the graph \cite{pozzi2013spread}.
Planarity ensures easy visualization of the network with the possibility to draw the network without edge-crossing.
Another appealing advantage of planar filtered networks concerns graphical modeling (e.g. Markov Random Fields \cite{kindermann1980markov}) where planarity (which limits the treewidth of the junction tree of the filtered graph) grants that some exact inference algorithms can be performed in an efficient fashion (see \cite{jaakkola2007approximate,bach2001thin}).
However, the algorithm so far proposed to construct the PMFG is computationally costly and cannot be applied to large datasets.
There is therefore  scope to search for novel algorithms that can construct planar filtered graphs in a computationally efficient way.
In the present paper we indeed introduce a computationally efficient  algorithm, the TMFG, that produces planar filtered graphs by optimizing an objective function (which we shall call ``score function'') by using local topological moves called $T_1$ and $T_2$ \cite{aste2012exploring}, the `Alexander move'  $A$ \cite{Alexander30}, in conjunction with a `vertex-swap' operator $S$. 

The TMFG algorithm has also the  advantage of allowing `online' updates of the planar graphs. 
Furthermore, the TMFG can be naturally applied to multipoint dependency measures associated with the 3- and 4-clique structure. 
When only $T_2$ and $S$ operators are used the algorithm  produces triangulated (or chordal) graphs together with their structure of cliques and separators. Chordal graphs have appealing properties: on the one side they are very well suited to modelling with Markov Random Fields (MRF) since they have a closed-form solution for the Maximum Likelihood Estimate (see~\cite{lauritzen1996graphical,wainwright2008graphical}) of the joint probability, on the other side chordal graphs are {\it perfect graphs} and as such are known to have polynomial time solutions for problems that are otherwise harder (e.g. graph coloring problem, maximum clique problem, and maximum independent set problem, see~\cite{grotschelgeometric}).
Graphical models with a chordal underlying graph have a particularly useful representation of the joint probability distribution and a closed form solution for the Maximum Likelihood Estimate (MLE). 
In the case of Gaussian graphical models the structure of the graph represents partial correlations between variables and the non-zero entries of the \emph{concentration} matrix (the inverse of the covariance matrix also called \emph{precision} matrix \cite{drton2009,lauritzen1984decomposable,lauritzen1996graphical}) coincide with the edges of the graph.
Additionally, the algorithm has the advantage that it is not restricted to planar topologies allowing higher-genus hyperbolic embeddings  to be explored \cite{AsteSherr,AsteSherr11,aste2012exploring,song2012building}.
Finally, given its local nature, the algorithm is ideally suited for parallelisation and GPU computing. 

This paper is organized as follows. In section \ref{s.1} we discuss some facts about Planar Maximally Filtered Graphs describing two algorithms used to generate such graphs; in section \ref{s.3} we introduce the TMFG algorithm and we highlight some characteristics of the algorithm that are relevant to applications with high-dimensional, frequently-updated datasets; in section   \ref{s.4} we offer some information-theoretic perspectives on the selection of a particular score function; finally in section \ref{s.results&discussion} we apply TMFG to several weight distributions showing that it is computationally  faster than PMFG achieving comparable or better results. 

\section{Planar Maximally Filtered Graphs} \label{s.1}

Algorithms for the extraction of  planar subgraphs from dense networks are relevant in several domains such as, for instance: 
i) the analysis of financial data -- where nodes generally represent financial assets (such as stock prices, spreads, liabilities, risk or liquidity indicators etc.) and the edges represent correlations or other measures of dependence between them (see \cite{aste2005complex,Tumminello26072005,pozzi2013spread,fiedor2014networks}); 
ii) facilities layout -- where  nodes represent the facilities and the edges the affinities between them (see \cite{foulds1976strategy,foulds1978graph,Liebers2001} for a survey); 
iii) integrated circuit design -- where nodes are the electrical elements and connections are the physical connections (see \cite{Lengauer:1990:CAI:92429});
iv) systems biology -- where nodes can represent proteins and edges protein interactions in a metabolic network (see \cite{song2007correlation});
v) social systems  -- where nodes represent social agents (e.g. individuals, companies, groups) and edges represent social interaction (see \cite{easley2010networks} for a detailed overview).
In some domains -- such as facility layout or integrated circuit design -- the constraint of planarity is a direct consequence of the two dimensional planar geometry of the problem, while in other domains, network planarity is a way to constraint the complexity of the graph reducing the degree of interwovenness \cite{aste2005complex}. 
Planarity is also desirable because many NP-hard problems have efficient polynomial-time solutions, or better approximations, for planar graphs (vertex coloring, edge coloring, independent vertex set, multicommodity flows, see \cite{nishizeki1988planar} for an introduction).

Let us here  briefly review two known algorithms that have been used to construct planar filtered networks from a matrix of weights: 
\begin{itemize}
\item PMFG \cite{Tumminello26072005}; 
\item Deltahedron heuristics and subsequent improvement \cite{foulds1978graph,Liebers2001,Osman2003}.
\end{itemize}
These algorithms provide estimates for optimal solutions to the \textsc{Maximum Weighted Planar Graph} problem.
Let us here recall that, given a complete edge-weighted graph $G(V,E)$, with vertex set $V$ and edge set $E$, the \textsc{Maximum Weighted Planar Graph} problem requires to build a planar subgraph $G'(V,E')$, with $E'\subset E$ such that adding another edge $e \in E \setminus E'$ would cause $G(V,E' \cup e)$ to be non planar and such that the sum of the weights is maximum (see \cite{Liebers2001} and \cite{Osman2003} for a detailed description of the problems and a survey). 

\subsection{Planar Maximally Filtered Graph}
The PMFG algorithm \cite{aste2005complex} searches for the maximum weighted planar subgraph by adding edges one by one (see \cite{MatlabPMFG}).
The resulting matrix is sparse, with $3(p-2)$ edges. 
The algorithm starts by sorting by weighting all the edges of a dense matrix of weights in non increasing order and tries to insert every edge in the PMFG in that order. 
Edges that violate the planarity constraint are discarded. 
The most computationally intense part of the algorithm is the planarity test, which is performed every time an edge insertion is attempted. 
It results that the PMFG construction performs an order of $p^2$  ($O(p^2)$) planarity tests on any dense $p\times p$ matrix of weights $W$. Assuming that the complexity for a planarity test in $O(p)$ (see ~\cite{hopcroft1974efficient, boyer2001simplified}) the computational complexity of the whole algorithm results in a $O(p^3)$  \cite{song2012hierarchical}. 

\subsection{Deltahedron heuristic }

\begin{center}
	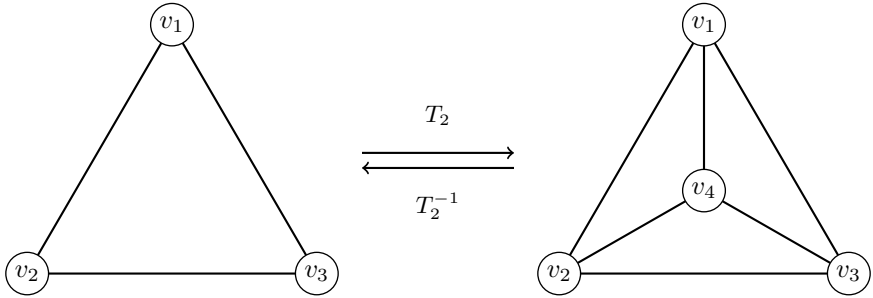
\begin{figure}[t]
		\begin{tikzpicture}
			\GraphInit[vstyle=Hasse]
			\SetVertexNormal[MinSize=16pt]
			\begin{scope}
				\grCycle[prefix=V,RA=2.2,rotation=90]{3}
				\AssignVertexLabel{V}{$v_1$,$v_2$,$v_3$}
			\end{scope}
			\draw (3.5,1) node {$T_{2}$};
			\draw[->, thick] (2.5,0.5) -- (4.5,0.5);
			\draw[<-, thick] (2.5,0.3) -- (4.5,0.3);
			\draw (3.5,-0.2) node {$T_{2}^{-1}$};
			\begin{scope}[xshift=7cm]
				\grTetrahedral[RA=2.2]
				\AssignVertexLabel{a}{$v_1$, $v_2$, $v_3$, $v_4$}
			\end{scope}
		\end{tikzpicture}
		\caption{$T_2$ move: addition of one vertex within a triangular face  \cite{AsteSherr,aste2012exploring,AsteSherr11,Dubertret98,Andrade05}. Its inverse, $T_2^{-1}$, removes a vertex  from inside a three-clique (in this case the clique $\left\lbrace v_1, v_2, v_3 \right\rbrace$).}
		\label{fig:deltah1}
	\end{figure}
\end{center}

The deltahedron heuristic \cite{foulds1978graph,Liebers2001} searches for approximate solutions of the WMPG problem starting from a tetrahedron, $K_4$, which is  planar. 
Then, at each step a vertex is added into a triangular face and three edges are added connecting the newly inserted vertex to the vertices of the triangular face.
This vertex insertion in a triangular face is called $T_2$ move (see Fig.\ref{fig:deltah1} and \cite{AsteSherr,aste2012exploring,AsteSherr11,Dubertret98,Andrade05}).
It is easy to see that the $T_2$ operator acts without breaking  planarity ensuring that the final network is planar. 
The triangular face is chosen in order to maximise the sum of the newly connected edges, while the vertices to be inserted are extracted from a pre-sorted list. The vertex list can be sorted according to two functions of the edge weights incident to the vertex, yielding two possible variants of the deltahedron heuristic:(i) the sum of the incident edge-weights or (ii) the maximum incident edge-weight. Different weightings lead to different ordering for the vertices and different results.

The deltahedron heuristic algorithm is not ``greedy'', for edge-insertion protocol, since the choice of the ordering of the vertices is done once at the beginning and there is no subsequent attempt at optimising the order of the vertices taking into account the local configuration and there is no known performance guarantee.  
However the algorithm is considerably faster than the PMFG, since every $T_2$ move keeps the planarity of the graph and therefore there is no need to test for planarity at each stage. 

An important feature of the graphs produced by $T_2$ moves is that they are chordal graphs: every cycle of length greater than 4 has a \emph{chord}, an edge not belonging to the cycle that joins two non-adjacent vertices. 
Chordal graphs are \emph{perfect} graphs and as such there are polynomial time algorithms for solving generally hard problems such as finding a maximum clique, graph coloring, and maximum independent set. 

In \cite{green1985heuristic, Osman2003} the deltahedron heuristic is improved by keeping a data structure of the most effective ways of inserting a vertex inside the faces (Green and Al-Hakim heuristic -- the GH-heuristic henceforth), essentially keeping a cache of the best and next-to-best options for inserting any of the remaining vertices. 
The cache is updated as the algorithm progresses. Optionally Osman et al. \cite{Osman2003} allow for a parameter that governs the``greediness'' of the algorithm. 
In section \ref{s.3} we will introduce a modified version of the GH-heuristic algorithm. 


\begin{center}
	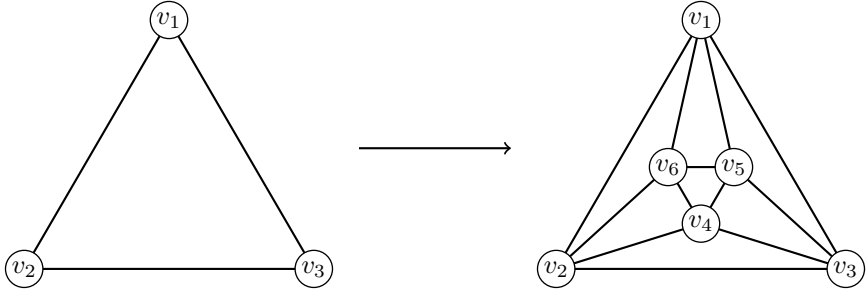
\begin{figure}[t]
		\begin{tikzpicture}
			\GraphInit[vstyle=Hasse]
			\SetVertexNormal[MinSize=14pt]
			\begin{scope}
				\grCycle[prefix=V,RA=2.2,rotation=90]{3}
				\AssignVertexLabel{V}{$v_1$,$v_2$,$v_3$}
			\end{scope}
			\draw[->, thick] (2.5,0.5) -- (4.5,0.5);
			\begin{scope}[xshift=7cm]
				\grCycle[prefix=v,RA=2.2,rotation=90]{3}
				\AssignVertexLabel{v}{$v_1$, $v_2$, $v_3$}
				\grCycle[prefix=a,RA=0.5,rotation=270]{3}
				\AssignVertexLabel{a}{$v_4$, $v_5$, $v_6$}
				\EdgeFromOneToSel{a}{v}{0}{1,2}
				\EdgeFromOneToSel{a}{v}{1}{0,2}
				\EdgeFromOneToSel{a}{v}{2}{0,1}
			\end{scope}
		\end{tikzpicture}
		\caption{Addition of three vertices in Leung's extension of the deltahedron heuristic.}
		\label{fig:deltah2}
	\end{figure}
\end{center}
 
\begin{center}
	\begin{figure}[t]
		\begin{tikzpicture}
			\GraphInit[vstyle=Hasse]
			\SetVertexNormal[MinSize=16pt]
			\begin{scope}
				\grCycle[prefix=V,RA=2.2,rotation=90]{4}
				\AssignVertexLabel{V}{$v_1$,$v_2$,$v_3$, $v_4$}
				\Edges(V0,V2)
			\end{scope}
			\draw (3.5,0.5) node {$T_{1}$};
			\draw[<->, thick] (3,0) -- (4,0);
			\begin{scope}[xshift=7cm]
				\grCycle[prefix=W,RA=2.2,rotation=90]{4}
				\AssignVertexLabel{W}{$v_1$, $v_2$, $v_3$, $v_4$}
				\Edges(W1,W3)
			\end{scope}
		\end{tikzpicture}
	\caption{$T_1$ move: rewiring of a shared edge between neighboring triangular faces.}
	\label{fig:T1}
	\end{figure}
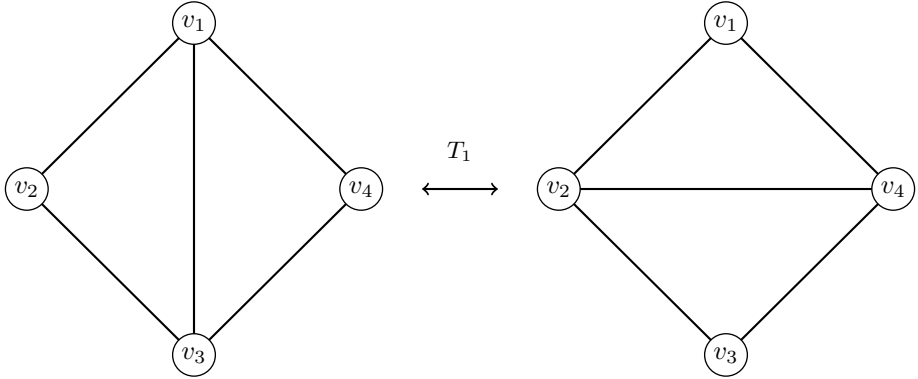
\end{center}

\begin{center}
	\begin{figure}[t]
		\begin{tikzpicture}
			\GraphInit[vstyle=Hasse]
			\SetVertexNormal[MinSize=14pt]
			\begin{scope}
				\grTetrahedral[RA=2.2]
				\AssignVertexLabel{a}{$v_1$, $v_2$, $v_3$, $v_4$}
			\end{scope}
			
			\draw (3.5,1) node {$T_2$};
			\draw[->, thick] (2.5,0.5) -- (4.5,0.5);
			
			\begin{scope}[xshift=7cm]
				\SetVertexNormal[MinSize=14pt]
				\grTetrahedral[RA=2.2]
				\AssignVertexLabel{a}{$v_1$, $v_2$, $v_3$, $v_4$}
			\end{scope}
			\begin{scope}[xshift=7.2cm]
				\SetVertexLabel
				\SetVertexNormal[MinSize=14pt]
				\Vertex[Math,x=0.35,y=0.35]{v_5}
				\Edge(a0)(v_5) \Edge(a2)(v_5) \Edge(a3)(v_5)
			\end{scope}
			
			\draw (-1,-3.) node {$T_2$};
			\draw[->, thick] (-2,-3.5) -- (0,-3.5);

			\draw (4,-3.) node {$T_1$};
			\draw[->, thick] (3,-3.5) -- (5,-3.5);

			\begin{scope}[xshift=1.2cm,yshift=-4cm]
				\grTetrahedral[RA=2.2]
				\AssignVertexLabel{a}{$v_1$, $v_2$, $v_3$, $v_4$}
				\SetVertexLabel
				\SetVertexNormal[MinSize=12pt]
				\Vertex[Math,x=0.47,y=0.4]{v_5}
				\Edge(a0)(v_5) \Edge(a2)(v_5) \Edge(a3)(v_5)
				\Vertex[Math,x=0.,y=-0.65]{v_6}
				\Edge(a1)(v_6) \Edge(a2)(v_6) \Edge(a3)(v_6)
			\end{scope}

			\begin{scope}[xshift=7.2cm, yshift=-4.0cm]
				\grCycle[RA=2.2, rotation=90]{3}
				\AssignVertexLabel{a}{$v_1$, $v_2$, $v_3$}
				\SetVertexLabel
				\Vertex[Math,x=-0.45,y=0.2]{v_4}
				\Vertex[Math,x=0.45,y=0.2]{v_5}
				\Vertex[Math,x=0.0,y=-0.5]{v_6}
				\Edge(a0)(v_5) \Edge(a2)(v_5) \Edge(v_4)(v_5)
				\Edge(a0)(v_4) \Edge(a1)(v_4) \Edge(a2)(v_5) 
				\Edge(a1)(v_6) \Edge(v_4)(v_6) \Edge(v_5)(v_6) \Edge(a2)(v_6)
			\end{scope}
			
		\end{tikzpicture}
		\caption{Demonstration that the Leung's extension in Fig.\ref{fig:deltah2} can be generated by using two $T_2$ and one $T_1$ moves.}
		\label{fig:xyz}
	\end{figure}
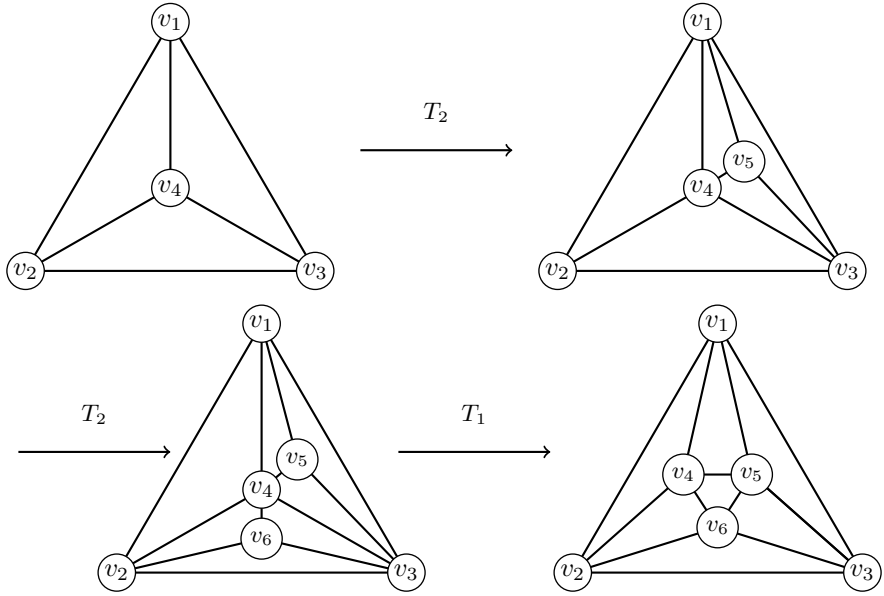
\end{center}

\begin{center}
	\begin{figure}[t]
		\begin{tikzpicture}
			\GraphInit[vstyle=Hasse]
			\SetVertexNormal[MinSize=16pt]
			\begin{scope}
				\grCycle[prefix=V,RA=2.2,rotation=90]{4}
				\AssignVertexLabel{V}{$v_1$,$v_2$,$v_3$, $v_4$}
				\Edges(V0,V2)
			\end{scope}
			\draw (3.5,0.6) node {$A$};
			\draw[->, thick] (3,0.3) -- (4,0.3);
			\draw[<-, thick] (3,-0.3) -- (4,-0.3);
			\draw (3.5,-0.6) node {$A^{-1}$};
			\begin{scope}[xshift=7cm]
				\grCycle[prefix=W,RA=2.2,rotation=90]{4}
				\AssignVertexLabel{W}{$v_1$, $v_2$, $v_3$, $v_4$}
				\SetVertexLabel
				\SetVertexNormal[MinSize=12pt]
				\Vertex[Math,x=0,y=0] {v_5}
				\Edges(W1,v_5)
				\Edges(W2,v_5)
				\Edges(W3,v_5)
				\Edges(W0,v_5)
			\end{scope}
		\end{tikzpicture}
	\caption{$A$ move: insertion of a vertex inside a plaquette made of two neighbouring triangular faces.}
	\label{fig:A}
	\end{figure}
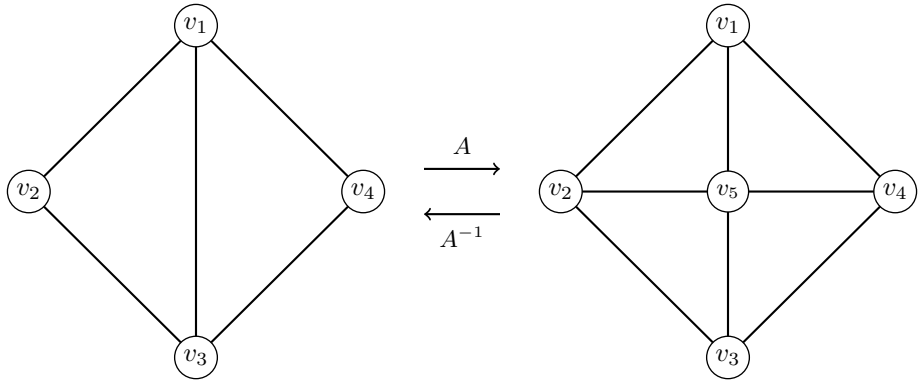
\end{center}

\subsection{Local topological moves: $\mathbf{T_1}$, $\mathbf{T_2}$, $\mathbf{A}, \enspace \& \enspace \mathbf{S}$} 
With the deltahedron heuristic we have already introduced the $T_2$ move that, as shown in Fig.\ref{fig:deltah1}, inserts vertex $v_4$ into the triangular face $\left\{v_1, v_2, v_3\right\}$  spliting it into three triangular faces $\left\{v_1, v_2, v_4\right\}$, $\left\{v_1, v_4, v_3\right\}$, and $\left\{v_4, v_2, v_3\right\}$. 
In the following we will call \emph{face} a three-clique that does not contain any vertex in the given embedding, reserving the word \emph{triangle} for a generic cycle of length 3.
We see that, after the $T_2$  move, $\left\{v_1, v_2, v_3\right\}$  is no longer a face but rather a 3-clique. 

In an extension of the deltahedron heuristic method, suggested by Leung \cite{leung1992new},  vertex insertion can happen either one vertex at a time (the $T_2$ move, as in Fig.\ref{fig:deltah1}) or three vertices at a time as in Fig.\ref{fig:deltah2}. 
This corresponds to the insertion of an octahedron within a triangular face \cite{song2012building}; clearly this is different from the  $T_2$ move that instead corresponds to the insertion of a tetrahedron.
However, such a move can be obtained by combining $T_2$ with another local move, called $T_1$ \cite{AsteSherr,aste2012exploring,AsteSherr11,Dubertret98}, consisting in switching neighbors among two adjacent triangles, as shown in Fig.\ref{fig:T1}.
In general, any  local  topological change of a surface triangulation that preserves embedding and results in a triangulation can be realized through the combination of the two elementary moves $T_1$ and $T_2$ \cite{Alexander30}.
However it should be pointed out that the application of $T_1$ could cause the graph to become no longer chordal.
For instance, the Leung extension (Fig.\ref{fig:deltah2})  can be produced via two $T_2$ and one $T_1$, as demonstrated in Fig.\ref{fig:xyz}.

Another move that we will use to build planar graphs is the $A$ move as described in Fig.\ref{fig:A}. Also in this case the move can be produced combining $T_1$ and $T_2$ and leads to non-chordal graphs. 

Finally, we will use the  `swap' operator, $S$, that re-labels sub sets of vertices of a graph as shown in Fig.\ref{fig:swap}, where it is acting on the vertices of a 4-simplex. 
This operation is trivial when the weights are identical, but will in general affect aggregate functions of the weights in a non-trivial way. 
The peculiarity of this operator is that it doesn't have to operate locally and it keeps topology unchanged preserving therefore planarity.  

In the following section we shall see how $T_2$, $T_1$, $A$ and $S$ moves can be used to generate planar filtered graphs as well as higher genus, non-planar, filtered graphs.

\begin{center}
	\begin{figure}[t]
		\begin{tikzpicture}
			\GraphInit[vstyle=Hasse]
			\SetVertexNormal[MinSize=16pt]
			\begin{scope}
				\grTetrahedral[RA=2.2]
				\AssignVertexLabel{a}{$v_1$, $v_2$, $v_3$, $v_4$}
			\end{scope}
			\draw (3.5,1) node {$S$};
			\draw[<->, thick] (2.5,0.5) -- (4.5,0.5);
			\begin{scope}[xshift=7cm]
				\grTetrahedral[RA=2.2]
				\AssignVertexLabel{a}{$v_1$, $v_4$, $v_2$, $v_3$}
			\end{scope}
		\end{tikzpicture}
		\caption{$S$ move: relabelling of the vertices of a 4-simplex. Note that the topology of the graph is unchanged.}
		\label{fig:swap}
	\end{figure}
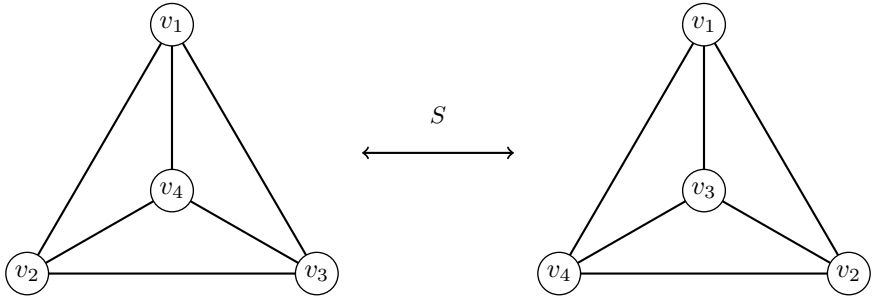
\end{center}

\section{Triangulated Maximally Filtered Graph} \label{s.3}

\subsection{TMFG construction}
TMFG algorithm starts from a clique of order 4 ($K_4$) and adds vertices by using the local move $T_2$. 
The novelty is that, at each step, the algorithm optimizes a {\it score function} (e.g. the sum of the weights of the edges). 
Similarly to the GH-heuristics, the method does not rely on any particular ordering of the vertices but, at every step, it calculates the score that would be obtained by adding any of the remaining vertices inside any feasible face. $T_2$ is applied to the vertex and face pair that leads to the maximum increase in score.
A naive implementation would require to evaluate the gain function for every pair consisting of a feasible vertex and a feasible face, thus resulting in an $O(p^2)$ calculations at every step and therefore $O(p^3)$ overall computational complexity. 
However, it is possible to maintain and update incrementally a cache with the information about the best possible pairing updating only the records affected by a move. Assuming the maximum requires $O(p)$ calculations for $p$ vertices, the overall number of calculations for the score functions is $O(p^2)$. 
This results in much faster computational times with respect to the $PMFG$. Differently  from \cite{Osman2003} we use a slightly different data structure to keep track of the vertices to insert into the feasible faces. We also keep track of the triangles that are no longer faces because this is relevant for subsequent modelling (see section \ref{s.4}).

Let us first focus on $T_2$ moves only.
After every application of $T_2$ the  cache is updated: some scores that were previously achievable are no longer feasible, while others become feasible and the corresponding score is calculated. 
More formally, we define a {\it score function} $S(v_h, \left\{v_a, v_b, v_c\right\})$ that quantifies the gain achievable by adding vertex $v_h$ inside the triangle $\left\{v_a, v_b, v_c\right\}$.

For instance, for a given, dense, matrix of weights $W$, the gain function can be the sum of the weights of the edges that will be added by inserting $v_h$ in face $\left\{v_a, v_b, v_c\right\}$: $S(v_h, \left\{v_a, v_b, v_c\right\}) = W(v_h, v_a)+ W(v_h, v_b)+ W(v_h, v_c)$.
In the next session we discuss an information theoretic interpretation of the score function. 

The cache is a structure made up of two vectors ($MaxGain$ and $BestVertex$) indexed by the faces in the planar graph present up to that point. 
Let us consider a given stage of the construction with $m$ triangular faces $t_i$, $i\in \left\lbrace 1, 2, \cdots, m \right\rbrace$ and $k$ remaining uninserted vertices ${v \in \left\lbrace v_1 \cdots v_k \right\rbrace }$.
The $MaxGain$ vector contains the value of the maximum gain over all remaining vertices for all triangular faces: 

\begin{equation}
\mathbf{MaxGain} = \left( \max\limits_{v \in \left\lbrace v_1 \cdots v_k \right\rbrace } S(v, t_1) , \max\limits_{v \in \left\lbrace v_1 \cdots v_k \right\rbrace } S(v, t_2) ,  \dots\max\limits_{v \in \left\lbrace v_1 \cdots v_k \right\rbrace} S(v, t_m) \right) \;\;.
\end{equation} \label{eq.maxgain}

The $BestVertex$ vector contains inside the list of vertices that attains the maximum gain for the specific triangular face:

\begin{equation}
\mathbf{BestVertex} = \left(\argmax\limits_{v \in \left\lbrace v_1 \cdots v_k \right\rbrace} S(v, t_1), \argmax\limits_{v \in \left\lbrace v_1 \cdots v_k \right\rbrace } S(v, t_2), \dots, \argmax\limits_{v \in \left\lbrace v_1 \cdots v_k \right\rbrace} S(v, t_m)\right)
\end{equation} \label{eq.bestvertex}

When  a vertex (say vertex $v_h$) is added to a certain triangular face (say face $t_j$) the two cache vectors must be updated by removing vertex $v_h$ from the list of remaining vertices, removing  face $t_j$ and adding three new faces. 
It is worth noting that $t_j$ becomes a \emph{clique separator} of the graph \cite{lauritzen1984decomposable}. 
The TMFG pseudocode is shown in Algorithm \ref{algo:TMFG}.  
For simplicity we have not given details of the application of the moves $T_1$,  $A$ and the swap operator $S$.
The TMFG generated by using $T_2$ only is a 4-clique tree.

\SetNlSkip{1em}
\SetInd{0.15em}{0.45em}
\begin{algorithm}[h] 
	\SetKwData{VertexList}{$\mathcal{V}$}
	\SetKwData{CurrentFaces}{$\mathcal{F}$}
	\SetKwData{Gains}{Gains}
	\SetKwData{MaxGain}{$MaxGain$}
	\SetKwData{BestVertex}{$BestVertex$}
	\SetKwData{W}{$\mathcal{W}$}
	\SetKwData{TMFG}{$\mathcal{P}$}
	\SetKwData{tetra}{$\mathcal{C}_1$} 	
	\SetKwData{ta}{$t_1$} 	
	\SetKwData{tb}{$t_2$} 	
	\SetKwData{tc}{$t_3$} 	
	\SetKwData{td}{$t_4$}
	\SetKwData{Triangles}{$\mathcal{T}$}
	\SetKwData{Separators}{$\mathcal{S}$}
	\SetKwData{Cliques}{$\mathcal{C}$}
	
	\SetKwInOut{Input}{input}\SetKwInOut{Output}{output}
	\Input{A dense $p \times p$ square matrix \W with positive weights (e.g. a matrix of squared correlation coefficients)}
	\Output{A sparse matrix, \TMFG, a filtered version of \W fulfilling the planarity constraint}
	
	$\tetra\leftarrow$ Tetrahedron, $\left\lbrace v_1, v_2, v_3, v_4 \right\rbrace$, with highest overall total score \;
	
	\tcp{Assign the four triangular faces in \tetra to the array \Triangles}
	$\Triangles\leftarrow \left\lbrace \left\lbrace v_1, v_2, v_3 \right\rbrace, 
									   \left\lbrace v_1, v_2, v_4 \right\rbrace,
									   \left\lbrace v_1, v_3, v_4 \right\rbrace,
									   \left\lbrace v_2, v_3, v_4 \right\rbrace
						  \right\rbrace$\;
	\tcp{Put the $p-4$ vertices not belonging to \tetra in the array \VertexList}
	$\VertexList\leftarrow \left\lbrace v_5, \cdots, v_p \right\rbrace$\;
	\tcp{Create an empty list of Separators}
	$\Separators\leftarrow \varnothing$ \;
	\tcp{Assign the first tetrahedron to the list of cliques}
	$\Cliques\leftarrow \tetra$\;
	$\TMFG\leftarrow \W(\tetra, \tetra)$\;
	Calculate \MaxGain for \Triangles and \VertexList as in Eq.(1) \;
	Calculate \BestVertex for \Triangles and \VertexList as in Eq. (2) \;
	\tcp{Insert $p-4$ vertices via $T_2$}
	\While{\VertexList is not empty}{
		Find the $t_i \in \Triangles$ and $v_i \in \VertexList$ that achieve the maximum in \MaxGain \;
		Insert $v_i$ into $t_i$ \tcp{this creates three new triangles $t_a$, $t_b$, $t_c$ }
		$\VertexList\leftarrow \VertexList \setminus v_i$\;
		$\Triangles\leftarrow \left(\Triangles \setminus \left\lbrace t_i \right\rbrace \right) \cup \left\lbrace t_a, t_b, t_c \right\rbrace$\;
		$\mathcal{S}_i \leftarrow \left\lbrace t_i \right\rbrace$\;
		$\Separators\leftarrow \Separators \cup \mathcal{S}_i$\;
		$\mathcal{C}_i \leftarrow \left\lbrace t_i, t_a, t_b, t_c \right\rbrace$\;
		$\Cliques\leftarrow \Cliques \cup \mathcal{C}_i$\;
		$\TMFG \leftarrow \TMFG + \W(\mathcal{C}_i, \mathcal{C}_i) - \W(\mathcal{S}_i,\mathcal{S}_i)$\;
		Update \MaxGain and \BestVertex to reflect the changes in \Triangles and \VertexList \;
	}
	\Return \TMFG\;
	\caption{TMFG algorithm} \label{algo:TMFG}
\end{algorithm}%

The TMFG algorithm can be extended to include $T_1$ and $A$ moves as well. 
In this case, the moves are local, internal to the plaquette made by  two joint triangles (i.e. $\{v_1,v_2,v_3\}$ and $\{v_2,v_3,v_4\}$ in Fig.\ref{fig:T1}).
The gain function for a $T_1$ move is associated to the removal of an edge (i.e. $(v_1,v_3)$ in Fig.\ref{fig:T1}) and the simultaneous addition of another edge (i.e. $(v_2,v_4)$ in Fig.\ref{fig:T1}). Similarly the gain for a move of type $A$ (as shown in Fig.\ref{fig:A}) results from the removal of one edge and the insertion of a new vertex and four new edges.
The use of $T_1$ and $A$ moves generally improve gain; however, we have verified that the algorithm with $T_2$ only produces very similar results.
Furthermore,  planar filtered graph with $T_1$ or $A$ moves are no longer clique trees but rather bubble-trees \cite{song2012hierarchical} which are in general no longer chordal. 
For instance see Fig.\ref{fig:xyz} where the application of $T_1$ creates a non-chordal graph: the cycle $v_1 - v_2 - v_6 - v_5 - v_1$ has length grater than 3 without internal chords. 
This can have some implications for dependency modeling, as we shall discuss in Section \ref{s.4}.
In the following we will therefore consider separately the two cases of TMFG constructed with and without $T_1$ and $A$.
In many cases the application of the swap operator $S$ results in higher overall gains. 
This operator has the advantage of leaving the overall topology unchanged but its use should be regulated by few local or heuristic criteria to avoid an increase in the complexity of the algorithm due to the increasing number of possible combinations. 
The case that we have implemented requires the evaluation of all the possible combinations of the four vertices involved in the execution of a $T_2$ operation. 
This requires some further changes to the cache vectors, but -- being applied locally -- it does not increase the overall computational complexity that remains $O(p^2)$.

The TMFG algorithm is not greedy with respect to edge insertion in the sense that the best possible move is chosen from a \emph{subset} of all the feasible edge insertions that preserve planarity. 
Nonetheless, we shall see that TMFG performs as well as -- or better than --  the PMFG for a large class of weight matrices, including squared correlation coefficient matrices from empirical time series which are relevant for modeling \cite{WolfPaper}.
 
\subsection{Dynamical adaptability}
Due to the local nature of the operators, $T_1$, $T_2$, $T_2^{-1}$, $A$, $A^{-1}$ and  (local)  $S$, used to construct the TMFG
one can continuously modify the network allowing  `online'  adaptability while new data are generated.
This is of practical importance because in real, big data, applications  information is changing dynamically with new data continuously fed causing changes in the matrix of weights that require modifications of the filtered graph.
Further, creation of new nodes is required when new elements/variables become relevant in the system.
Conversely, elements/variables can eventually become irrelevant and the corresponding vertices should be eliminated from the graph.
The implementation of these moves requires keeping a cache matrix of gains continuously updated and dynamically checking for  moves that improve total gains.

\subsection{Parallelization and big data}
The local nature of TMFG construction and dynamical adaptation through $T_1$, $T_2$, $T_2^{-1}$, $A$, $A^{-1}$ and  (local)  $S$, moves make it ideal for parallelization. 
There are several possibilities for parallelization and it is beyond the purpose of the present work to implement a parallel algorithm for TMFG. 
Let us however discuss briefly a possible parallel implementation of the TMFG.
One of the main features of planar triangulations is that three-cliques uniquely divide the network into two  `inside' and `outside' subgraphs within a nested hierarchical structure \cite{song2011nested}.
This means that, given a seed structure of three-cliques, each clique can develop its inside subgraph independently.
A processor can be assigned to each seed clique and calculations can be performed locally. 
Given that each separating clique divides roughly the graph into two parts one can compute the TMFG in $O(p)$ using $O(\log p)$ processors. Another issue related to big data is the size of the score vectors. It is clear from the construction that the size of the cache grows linearly with the dimension of the problem and that triangles in the basis can be assigned to different processors, allowing parallel updates of the cache.

\subsection{Memory usage}

In the case of pair-wise dependence (such as correlation) both the deltahedron heuristic and the PMFG require to compute in advance the entire correlation matrix, while the TMFG does not use the full information from the correlation matrix and could calculate only the correlations necessary for the incremental update of the gain vectors. This is an advantage already for correlation measures, in the (numerous) cases where the number of observations ($q$) is less than the number of variables ($p$): in fact it could require much less memory (approximately $p \times q$) to store the time series of the observations and calculate the correlations on-demand, rather than calculating and storing a large correlation matrix (approximately $\frac{p \times (p-1)}{2}$). This fact is even more relevant for multi-point dependencies (e.g. partial correlation, mutual information, ...): in these cases the TMFG would still require only to store the time series in memory and would require the calculation of the relevant gain functions only, while other methods would require the storage of large amounts of data (e.g. order of $p^3$ for a three-points dependency measure).

\section{Modeling with TMFG: information theoretic perspective }\label{s.4}
In complex systems, such as financial markets, a large number of interdependent variables are typically involved.
The TMFG is a way of filtering the structure of interrelation between the variables reducing it to a network of most relevant  interactions.

Modeling the system statistically  consists in identifying the joint probability distribution that best describes the observed collective behavior of the variables.

Specifically, given a set of observations $\{x_1(1),...,x_1(q)\}$, $\{x_2(1),...,x_2(q)\}$, ... $\{x_p(1),...,x_p(q)\}$  of $p$ random variables ${\mathbf X} = \{X_1,X_2,...,X_p\}$, one aims to estimate a joint probability distribution function $Q({\mathbf X})$ that is the best representation of the  `true' multivariate probability distribution function $P({\mathbf X})$ from which the set of observations are drawn.
Clearly, $P({\mathbf X})$ is unknown and the only information available are the observations  $\{x_1(1),...,x_1(q)\}$, $\{x_2(1),...,x_2(q)\}$, ... $\{x_p(1),...,x_p(q)\}$  from which $Q({\mathbf X})$ must be estimated.

Information filtering graphs can be used to compute $Q({\mathbf X})$. 
The main advantage is that these graphs are locally low dimensional (e.g. the largest clique is $K_4$ when planarity is enforced) which makes sampling tractable also with limited amount of data \cite{WolfPaper}. 
Further, the TMFG constructed from $T_2$ moves results in a tree made of 4-cliques separated by 3-cliques (called separators in the following). 
This is a particular case of a \emph{triangulated} or \emph{chordal} or  \emph{decomposable} graph \cite{lauritzen1984decomposable}.
From the theory of graphical models (see \cite{lauritzen1996graphical}) we know that the probability distribution function associated to a decomposable graph -- such as the TMFG --  admits the representation:
\begin{equation} \label{eq:graphmodelQ}
Q({\mathbf X})= \frac{\prod_{c \in Cliques}\phi_c({\mathbf X}_c)}{\prod_{s \in Separators}\phi_s({\mathbf X}_s)} \;.
\end{equation}

Where $\phi_c({\mathbf X_c})$ and $\phi_s({\mathbf X_s})$ are the marginal probabilities  of the sub sets of variables associated respectively with the 4-clique $X_c$ and separator $X_s$.

 Eq.\ref{eq:graphmodelQ} reduces the $p$-dimensional problem of estimating the joint probability distribution function $Q({\mathbf X})$ to the estimation of a set of 3- and 4-dimensional local marginal probabilities $\phi_s({\mathbf X_s})$ and $\phi_c({\mathbf X_c})$.
Such a dimensionally reduction helps greatly in the estimation of the joint probability.
The open question is now to measure how well the, unknown, true joint distribution  $P({\mathbf X})$ is represented by the model estimation $Q({\mathbf X})$ factorized over the TMFG.
To this end we can measure the dissimilarity between the two probability density functions which is given by the Kullback-Leibler divergence \cite{kullback1951}:
\begin{equation} \label{eq:KL}
D_{KL}(P \parallel Q) = \sum_{{\mathbf X}} P({\mathbf X}) \log \left(\frac{P({\mathbf X})}{Q({\mathbf X})}\right)\;;
\end{equation}

For simplicity we are considering discrete variables, a similar treatment can be developed for continuous variables.
The goal is to construct the TMFG that minimizes such a distance.
By substituting Eq.\ref{eq:graphmodelQ} into Eq.\ref{eq:KL} we obtain
\begin{eqnarray} \label{eq:KLfactors}
D_{KL}(P \parallel Q) =
 \sum_{{\mathbf X}} P({\mathbf X}) \log \left(P({\mathbf X})\right) \\ \nonumber
- \sum_{c \in Cliques}  \sum_{{\mathbf X_c}} \phi_c({\mathbf X_c}) \log \left(\phi_c({\mathbf X_c})\right) \\ \nonumber
+ \sum_{s \in Separators}  \sum_{{\mathbf X_s}} \phi_s({\mathbf X_s}) \log \left(\phi_s({\mathbf X_s})\right)  \;\;.
\end{eqnarray}

From a information theoretic perspective the first term in Eq.\ref{eq:KLfactors} (with a minus sign),
\begin{equation}
H =- \sum_{{\mathbf X}} P({\mathbf X}) \log \left(P({\mathbf X})\right)\;,
\end{equation}
quantifies the total amount of uncertainty in the system, measuring the number of bytes (if base-2 logarithms are used) necessary to define a state. 
The other two terms in Eq.\ref{eq:KLfactors}:
\begin{equation}\label{eq:Hm}
H_m = - \sum_{c \in Cliques}  \sum_{{\mathbf X_c}} \phi_c({\mathbf X_c}) \log \left(\phi_c({\mathbf X_c})\right) 
+ \sum_{s \in Separators}  \sum_{{\mathbf X_s}} \phi_s({\mathbf X_s}) \log \left(\phi_s({\mathbf X_s})\right) 
 \end{equation} 
 also quantify an uncertainty, but in this case, associated with the model of the system.
In other words, by adopting the TMFG structure of interactions,   $H_m$ measures the number of bytes necessary to define the state of the system when only the interrelations among variables associated with edges in the TMFG are considered. 

An algorithm to construct the TMFG with the aim of minimizing $D_{KL}(P \parallel Q)$ can be implemented by choosing at every stage the move that minimally increases $H_m$ consistently with all other constraints.
In particular, considering the TMFG construction via $T_2$ moves,  the contribution to $D_{KL}(P \parallel Q)$ from the insertion of a vertex $v$ added inside an existing triangular face $t$ generating a 4-clique $u$ is:
\begin{equation} \label{eq:GainIT}
S(v,t)  =   \sum_{{\mathbf X}_u} \phi_u({\mathbf X}_u) \log \left(\phi_u({\mathbf X}_u)\right) 
-  \sum_{{\mathbf X}_t} \phi_t({\mathbf X}_t) \log \left(\phi_t({\mathbf X}_t)\right)  \;\;.
\end{equation}
From an information theoretic perspective $-S$ is the amount of uncertainty introduced in the model by including a variable $v$, the TMFG structure should be constructed in a way to minimize such uncertainty. 
In \cite{WolfPaper} the case for normal multivariate distributions is discussed in details.

\section{Examples of TMFG construction and comparison with PMFG} \label{s.results&discussion}

A vast literature has demonstrated that PMFG can be used to retrieve meaningful information about the structure of interdependency in complex datasets \cite{Tumminello26072005,song2012hierarchical,musmeci2014relation,musmeci2014risk,pozzi2013spread}, it is therefore natural to compare the performances of the TMFG with the ones of the PMFG.

\begin{figure}[h]
\centering
\includegraphics[scale=0.6,width=13cm]{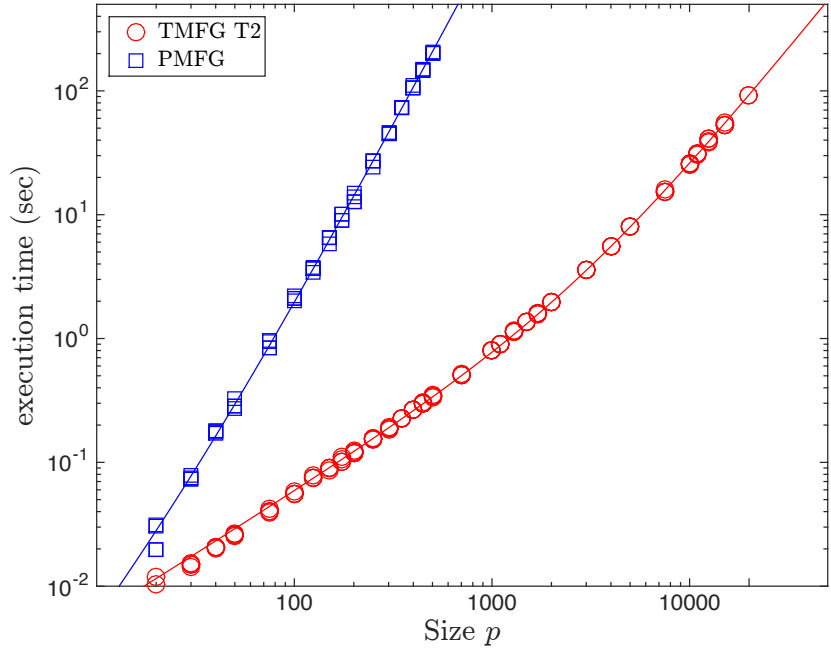} 
\caption{Demonstration that TMFG is faster and scalable with respect to the PMFG.
Comparison between execution times for TMFG and PMFG for different values of $p$  ranging between 50 and 10000. 
Lines are the 2-parameters best polynomial fits (see text).
}
\label{fig:executionTimes}
\end{figure}

Let us first look at the scaling of execution times for TMFG and PMFG algorithms as function of the size $p$ of the  weight matrix $W$; results are reported in  Fig.\ref{fig:executionTimes} (seconds on a 2.6 GHz Intel Core i7\textregistered). 
We  observe  that TMFG execution times scale with the matrix dimension size $p$ approximately as $O(p^2)$  while  PMFG scales approximately as $O(p^3)$. 
The 2-parameters best polynomial fits give respectively: $T_{TMFG} \sim 2 \cdot 10^{-7} \cdot p^{2} + 6 \cdot 10^{-4} p$  and 
$T_{PMFG} \sim 2 \cdot 10^{-6} \cdot p^{3} + 3 \cdot 10^{-5} p^2$.
Overall we can see that execution times are several orders of magnitude faster for TMFG than PMFG.

We have then compared the total retained edge weight for the following four variants of the TMFG construction:
\begin{enumerate}
\item TMFG: the base version of the algorithm. It uses only $T_2$ operators. This version of the algorithm produces chordal graphs.
\item TMFG-T1: uses $T_2$ followed by an optimisation stage where a number of $T_1$ moves are performed after every insertion of a new vertex. 
\item TMFG-S: a variant of the basic algorithm with $T_2$ followed by local optimization with $S$. 
\item TMFG-A: a variant of the algorithm with $T_2$ followed by local optimization with  $T_1$ and $A$.
\end{enumerate}

We have tested 9 types of random weight matrices $W$ with different weight distributions:

\begin{enumerate}
\item Beta distribution with shape parameters $\alpha=0.5$ and $\beta=3$. This distribution is heavily skewed and is characterised by very low density on the right side of the interval $ [0, 1] $. 
\item Beta distribution with shape parameters $\alpha=3$ and $\beta=0.5$. This distribution is skewed in the opposite direction and has a high density near the right extreme of the interval $ [0, 1] $. 
\item Pareto distribution with power law exponent equal to 1. This distribution has a fat tail.  
\item Pareto distribution with power law exponent equal to 2. This distribution has still a fat tail, but thinner than the previous one.   
\item Random matrix of correlations of 400 time series generated by simulating 20 normally distributed common factors.   
\item Random matrix of correlations of 400 time series generated by simulating 50 common factors. This matrix shows less structure than the one generated using 20 factors.  
\item Random matrix of correlations of 400 time series generated by simulating 100 common factors. This matrix shows less structure than the two above.  
\item Uniform distribution over $[0, 1]$.  
\item Square of a real correlation matrix coefficients computed from daily log-returns of 342 US stocks, across a period of 15 years (form  Jan 1997 to Jul 2012) (see \cite{musmeci2014relation}). 
\end{enumerate}
All matrices are symmetric and have size $p=400$ except the real correlation data that have sizes $p=342$.
For all the weight matrices (excepting for the real correlation matrix) we have compared results for 100 samples.
For the real correlation matrices we generated matrices by random sampling the starting point of 100 time windows of length 1000 data points over a period of 4500 points in total. 
Table \ref{t.perf} reports the average relative performance, defined as the ration between the sum of the edge weight in the the four variants of TMFG with respect to the sum of the edge weight in the PMFG. 
It shows that the PMFG is usually more effective when the density of high weights is low, while the TMFG is more effective when the density of high weights is higher or limited. 
This result is to be expected since the PMFG is less constrained than the TMFG in picking up isolated high-weight edges one at a time, while the TMFG is more efficient in selecting subsets of edges with a high total sum. For the random matrices of correlation we see that the TMFG performs better than the PMFG in filtering the more structured matrix generated using 20 factors. In the real case we see that the TMFG is marginally better than the PMFG.
We conclude that TMFG is in  general performing comparably well, and sometimes better than the PMFG.
We observe that the TMFG tends to improve relative performance as the size of the matrix increases (see Table \ref{t.increasing_perf}).
When other moves are used TMFG improves performances with best performances obtained by the TMFG-A variant.

\begin{table} \footnotesize \centering
\begin{tabular}{|p{2.5cm}|m{1.7cm}|m{1.7cm}|m{1.7cm}|m{1.7cm}|m{2.5cm}|} 
\hline 
\mbox{Weight matrix} coefficients distribution & {TMFG/ PMFG }& TMFG-T1/ PMFG & TMFG-S/ PMFG & TMFG-A/ PMFG  & TMFG (Time)/ PMFG (Time)\\ 
\hline 
Beta(0.5,3) & 95.42\% & 96.24\% & 95.72\% & 99.89\% & 0.16\% \\ 
\hline 
Beta(3, 0.5) & 104.70\% & 104.73\%	& 104.77\% & 104.80\% & 0.14\% \\ 
\hline 
Pareto(1) & 99.97\% & 99.97\% & 99.97\% & 99.97\% & 0.17\% \\  
\hline 
Pareto(2) & 97.94\%	& 98.00\% &	98.02\% & 98.32\% & 0.17\% \\ 
\hline 
Random Matrix (20 factors) & 102.23\% &	102.63\% & 102.57\% & 103.77\% & 0.22\% \\ 
\hline 
Random Matrix (50 factors) & 100.30\% & 100.82\% & 100.64\% & 102.54\% &  0.21\%\\ 
\hline 
Random Matrix (100 factors) & 98.46\% & 99.14\% & 98.86\% & 101.42 \% &  0.21\%\\ 
\hline 
Uniform & 116.27\% &	116.29\%	& 116.34\% & 116.89\% & 0.15\% \\ 
\hline
Real correlation matrix & 100.11\%	& 100.17\%	 & 100.24\% & 100.42\% &  0.15\% \\ 
\hline 
\end{tabular}
\caption[Table \ref{t.perf}]{ \label{t.perf}  Average relative performances (ratio between sum of edge weights) of the TMFG algorithm with respect to the PMFG. 
Four TMFG variants and nine different weight distributions.  
Note that TMFG and TMFG-S are chordal graphs. }
\end{table}

\begin{table}[h] \footnotesize \centering
\begin{tabular}{|p{2.5cm}|m{1.7cm}|m{1.7cm}|m{1.7cm}|m{1.7cm}|}
\hline 
 Weight matrix size $p$ & TMFG /PMFG & TMFG-T1 /PMFG & TMFG-S /PMFG & TMFG-A /PMFG   \\ \hline
50&88.68\%&88.82\%&89.35\%&95.93\%\\ \hline
100&90.49\%&93.13\%&92.31\%&98.14\%\\ \hline
150&92.14\%&93.77\%&90.44\%&95.26\%\\ \hline
300&93.73\%&95.6\%&94.63\%&100.06\%\\ \hline
500&96.36\%&96.79\%&96.6\%&100.98\%\\ \hline
700&98.83\%&100.49\%&98.92\%&103.58\%\\ \hline
850&98.93\%&99.95\%&99.99\%&103.83\%\\ \hline
1000&100.33\%&100.56\%&100.71\%&105.39\%\\ \hline
1200&101.34\%&102.16\%&101.16\%&105.24\%\\ \hline
\end{tabular}  \caption[Table \ref{t.inc_perf}]{\label{t.inc_perf} \label{t.increasing_perf}
 Example of relative increase in performance of the TMFG algorithm with respect to PMFG when dimensionality $p$ increases. 
The underlying distribution is a Beta(0.5, 3).}
\end{table}

%

\section{Conclusions}

We have described a new family of algorithms, TMFG, to retrieve approximate solutions to the \textsc{Maximal Planar Graph} problem, which have the following desirable characteristics:

\begin{enumerate}
\item  TMFG is faster than  PMFG with execution times increase as the square of the weight matrix size $p$, while PMFG increases with the third power of $p$.
\item  TMFG constructed with $T_2$ and $S$ only produces chordal graphs. This opens the door to the use of Markov Random Fields as a modelling tool. The fact that the maximum dimension of cliques is controlled by the topology entails that efficient inference algorithms can be used.
\item  From TMFG construction the structure of cliques and separators is automatically retrieved.
\item The local nature of the TMFG construction  allows one to model dependence using genuinely multivariate  score functions over the clique elements such as Mutual Information, Total Correlation, Partial Correlation, Likelihood functions etc. (as opposed to bivariate functions such as correlation).
\item  TMFG does not require  preliminary calculation and sorting of all the values of the score function. In cases when the distribution is based on 3 or more variables this constraint is severe.
\item In cases where $ p \gg q $ the memory footprint of the algorithm can be reduced by keeping the $ p\times q $ data points in memory and calculating the dependence function on-demand, instead of the $p^2$ values of a correlation matrix or the $ p^d $ values of a $d-$variate dependence function.
\item The use of local and non-local operators ($T_1, T_2, A, S$) allows one to change the topology of the network in an easy and controlled way as the network evolves. 
\item The geometric formulation of the algorithm allows one to consider generalisation higher genus to simplicial complexes structures. 
\item The limited treewidth of the filtered network makes it amenable to efficient (and often exact) inference \cite{wainwright2008graphical}.
\end{enumerate}

Future developments of the TMFG method are in the direction of building networks with a richer structure beyond and also below planarity while keeping a controlled complexity. 
We intend to apply this filtered networks to sparse modeling with application to physical, financial and biological systems.
We intend to develop applications of this filtered network construction and efficient inference and use this tool to identify large scale features of financial networks in different regimes and to apply the  inference algorithms  to a number of problems in financial risk management, such as risk aggregation and allocation, simulation, stress testing and incorporation of non-homogeneous sources of information (such as asset prices, macroeconomic variables, and also expert opinion) into a risk management framework.

\textbf{Acknowledgements} 
The authors acknowledge very useful discussions with Simone Severini.
TA acknowledges support of the UK Economic and Social Research Council (ESRC) in funding the Systemic Risk Centre [ES/K002309/1].
TDM wishes to thank the COST Action TD1210 for partially supporting this work. 
GPM thanks Alun Wyn-jones for reviewing previous drafts.
The authors thank Wolfram Barfu{\ss} for many useful discussions and suggestions.

\bibliographystyle{unsrt}

\end{document}